# THE PROBLEM OF FRICTION IN TWO-DIMENSIONAL RELATIVE MOTION

Dariusz Grech
Institute of Theoretical Physics, University of Wroclaw,
pl. Maksa Borna 9, 50-205 Wrocław, Poland
e-mail: dgrech@ift.uni.wroc.pl

Zygmunt Mazur
Institute of Experimental Physics, University of Wroclaw,
pl. Maksa Borna 9, 50-205 Wrocław, Poland
e-mail: zmazur@ifd.uni.wroc.pl

ABSTRACT:
We analyse a mechanical system in two-dimensional relative motion with friction. Although the system is simple, the peculiar interplay between two kinetic friction forces and gravity leads to the wide range of admissible solutions exceeding most intuitive expectations. In particular, the strong qualitative dependence between behaviour of the system, boundary conditions and parameters involved in its description is emphasised. The problem is intended to be discussed in theoretical framework and might be of interest for physics and mechanics students as well as for physics teachers.

## 1. INTRODUCTION

Friction forces make a part of elementary course of dynamics. Usually one states that the direction of the kinetic friction force acting on a given object is opposite to the direction of its motion and the magnitude of friction force is given as $F = fN$, where N is a normal reaction force acting on the object and $f$ is the coefficient of kinetic friction. Authors of physics textbooks very seldom emphasize that the direction of friction is always antiparallel to the *relative* velocity of two rubbing surfaces. Almost all examples of motion with friction we meet in college and even in university physics textbooks do not show the importance of the above-mentioned fact.

This paper describes behaviour of a simple mechanical system which should draw students' attention to the issue of relative motion of surfaces with friction.

Let us consider a mechanical system shown in details in Fig. 1. A block *B* of mass *m* is put on the plane $\sigma$ moving with constant velocity $\vec{u}$ parallel to its surface (like a belt conveyor). The coefficient of kinetic friction between the block and the surface of plane is given as $f_1 > 0$. An additional vertical rough wall *W* is put across and slightly above the moving plane to forbid the motion of the block in $\vec{u}$ direction. The surface of plane is inclined to the horizontal with an angle $0 < \alpha < \pi/2$ in such a way that block *B* can slide down along the wall W (see Fig. 1). The coefficient of kinetic friction between the wall *W* and the block *B* is assumed $f_2 > 0$. The whole system is in the uniform, vertical gravity field $\vec{g}$, which is the source of motion for *B* against the wall.

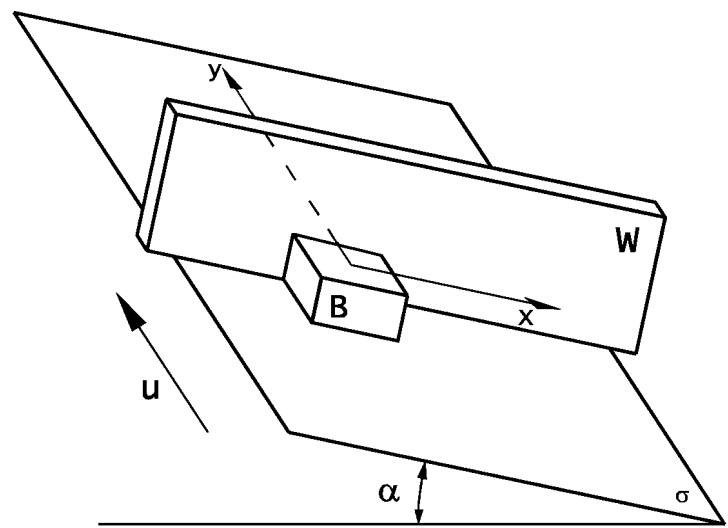

**Figure 1.** A view of the system.

Most students and even experienced physicists asked about how this system will behave answer instinctively without calculations that *B* will constantly accelerate if it is only able to start to move on its own, due to gravity, at some initial moment $t = 0$. Such expectation seems to be very natural but is it really the case here? We will show this expectation is completely unjustified and that the real behaviour of our mechanical system may, in general, seriously contradict the common sense point of view. Moreover, the particular choice of values for $f_1$, $f_2$, and $\alpha$ parameters may be the source of chaotic motion for *B*.

The essential role in proper description of this problem is played by the kinetic friction force [1]. As we already mentioned its direction is always antiparallel to the *relative* velocity of two rubbing surfaces*. The interplay between two kinetic friction forces present in the system, and described below, leads to amazing motion of the block B we analyse theoretically in this paper.

## 2. THE DYNAMICAL SYSTEM AND ITS EQUATION OF MOTION

Let us introduce the system of coordinates OXY in such a way that OY axis is perpendicular to the wall $W$ and OX axis lies in the plane $\sigma$ (see Fig.1). The relative velocity $\vec{v}_{rel}$ of $B$ against the plane is shown in Fig.2a.

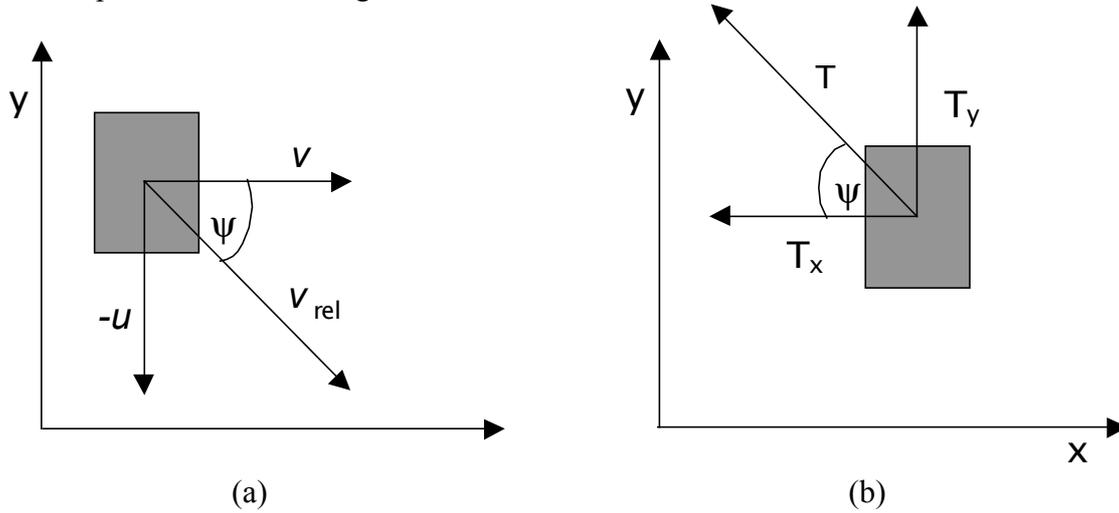

(a)          (b)

**Figure 2 (a)** The relative velocity of the block B against the plane σ **(b)** The decomposition of kinetic friction force $\vec{T}$ between the block B and the plane σ in OXY system of coordinates.

Thus we have:
$$\vec{v}_{rel} = [v, -u] \quad (1)$$
where $v$ is the linear velocity of the block along W.

Following this consideration we find components of kinetic friction force $\vec{T}$ between $B$ and $\sigma$ (see Fig. 2b):
$$\vec{T} = [T_x, T_y] \quad (2)$$
where:
$$T_x = mgf_1 \cos\psi \cos\alpha \quad (3)$$
$$T_y = mgf_1 \sin\psi \cos\alpha \quad (4)$$

and
$$\cos\psi \stackrel{df}{=} \frac{v}{\left(v^2 + u^2\right)^{1/2}} \quad (5)$$

---

* In some physics textbooks (see e.g.[2]) the direction of kinetic friction force is simply defined as antiparallel to the velocity of moving object. Such definition is obviously incorrect because it does not make Newton laws of dynamics invariant for various inertial systems of reference. Indeed, if we change the reference system, the velocity (its direction and magnitude) also changes. However, kinetic friction force and any other force involved must remain unchanged. Otherwise, the second Newton's law would be obviously broken.

From Newton's laws of dynamics the equations of motion for B clearly read:
$$\begin{cases} ma = mg \sin\alpha - T_x - f_2 N \\ \phantom{ma = }N = T_y \end{cases} \quad (6)$$

where $a = \dfrac{dv}{dt}$ is the acceleration of B along OX axis and $N$ – the magnitude of reaction force of the wall $W$ against B.

With the use of relations (3)-(5) we get from (6) the final expression for block acceleration in OX direction:

$$\frac{dv}{dt} = g \sin\alpha \left(1 - \frac{x_1 v(t) + x_2 u}{\left(v^2(t) + u^2\right)^{1/2}}\right), \quad (7)$$

where we introduced new parameters to be used from now on:

$$x_1 \stackrel{df}{=} s f_1 \quad (8)$$

$$x_2 \stackrel{df}{=} s f_1 f_2 \quad (9)$$

$$s \stackrel{df}{=} \operatorname{ctg} \alpha \quad (10)$$

It is worth to notice from (7) that if the initial velocity of B is $v_0 \equiv v(t=0) = 0$ then the block is able to start to move (i.e. $a(t=0) > 0$) only when $x_2 < 1$.

Contrary, the block rests all the time against the wall only if $x_2 \geq 1$. Indeed, in this case we have from (6):

$$mg \sin\alpha = F_T \quad (11)$$

where $F_T$ is the static friction force between B and W limited by its maximal value:

$$F_T \leq f_2 T_y \quad (12)$$

Combination of Eqs(11), (12) and (4) provides the required condition $x_2 \geq 1$.

It is interesting to look at the description of motion of this system in matrix form what we do in Appendix.

## 3. DISCUSSION OF SOLUTIONS

The subject of further analysis is the non-linear differential equation given in Eq. (7). The detailed analytical solution of this equation is troublesome and it is much more convenient to solve it numerically. However, basic properties of admissible solutions can be found very easy at elementary mathematics level. In this paper we will additionally use the simple Euler's method [3] which is stable for the considered case to illustrate qualitatively these solutions.

We see from the form of Eq.(7) that $\dfrac{dv}{dt} > 0$ if the right-hand side (RHS) of Eq.(7) satisfies the condition:

$$\left(v^2(t) + u^2\right)^{1/2} > x_1 v(t) + x_2 u \quad (13)$$

For positive $v(t)$, $u$, $x_1$, $x_2$, it is equivalent to the square inequality with respect to $v(t)$:

$$v^2(t)\left(1 - x_1^2\right) - 2u x_1 x_2 v(t) + u^2\left(1 - x_2^2\right) > 0 \quad (14)$$

with the discriminant:

$$\Delta = 4u^2\left(x_1^2 + x_2^2 - 1\right) \quad (15)$$

Let us first discuss the case $x_2<1$. It is already seen that the list of admissible solutions in this case may significantly exceed our naive expectations. Here we meet three possibilities for $x_1$ to be analysed: $x_1>1$, $x_1<1$ and $x_1=1$. The RHS of Eq.(7) versus $v(t)$ is plotted for these possibilities in Fig. 3a. We discuss them in following subsections.

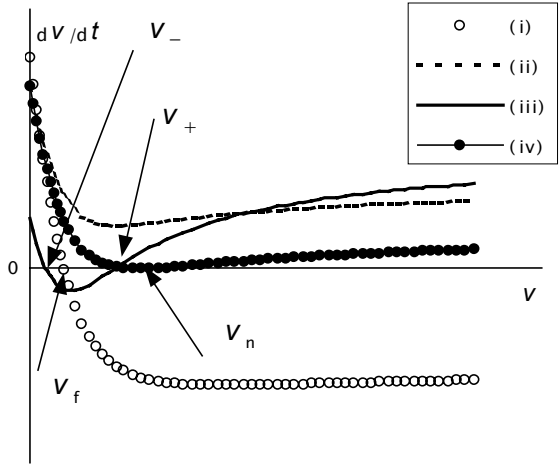
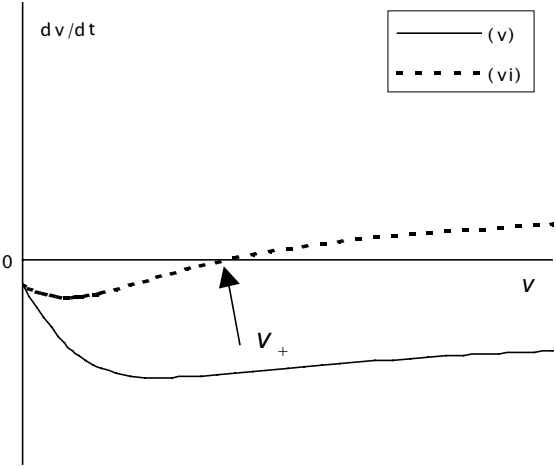

**Figure 3a.** The acceleration of the block B versus velocity $v$ for $x_2<1$:
  (i) $x_1 \geq 1$,
  (ii) $x_1<1$ and $x_1^2 + x_2^2 < 1$,
  (iii) $x_1<1$ and $x_1^2 + x_2^2 > 1$,
  (iv) $x_1<1$ and $x_1^2 + x_2^2 = 1$

**Figure 3b.** The acceleration of the block B versus velocity $v$ for :
  (v) $x_2 \geq 1$ and $x_1 \geq 1$,
  (vi) $x_2 \geq 1$ and $x_1 < 1$

### 3.1. SOLUTIONS FOR $X_1 > 1$

The range of $(x_1, x_2)$ parameters involved in these solutions is shown in Fig. 4a.
From Eq.(15) we have $\Delta > 0$ and therefore two distinct real roots of inequality (13) (and also those of RHS of Eq.(7)) exist:

$$v_\pm = u \frac{x_1 x_2 \pm \left(x_1^2 + x_2^2 - 1\right)^{1/2}}{1 - x_1^2} \qquad (16)$$

Clearly $v_+ < 0$, while $v_- > 0$. The latter relation follows from the inequality:

$$x_1^2 - 1 > x_2^2 \left(x_1^2 - 1\right) \qquad (17)$$

obviously satisfied for $x_1 > 1$ and $x_2 < 1$, which can also be written as:

$$\left(x_1^2 + x_2^2 - 1\right)^{1/2} > x_1 x_2 \qquad (18)$$

Hence we have:

$$\begin{cases} \dfrac{dv}{dt} > 0, & \text{if} \quad 0 \leq v \leq v_f \\ \dfrac{dv}{dt} = 0, & \text{if} \quad v = v_f \\ \dfrac{dv}{dt} < 0, & \text{if} \quad v > v_f \end{cases} \qquad (19)$$

where

$$v_f \stackrel{df}{=} v_- = u \frac{\left(x_1^2 + x_2^2 - 1\right)^{1/2} - x_1 x_2}{x_1^2 - 1} \qquad (20)$$

It means that $v(t) = v_f$ is the stationary and stable solution of Eq.(7) [4]. The numerical simulation of integral curves with various boundary conditions for considered range of $x_1$, $x_2$ parameters is shown in Fig.4b.

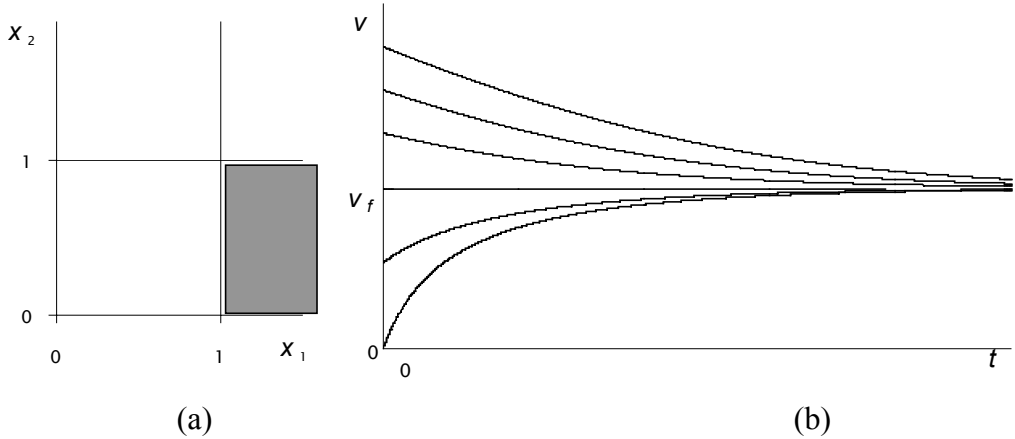

(a)          (b)

**Figure 4. (a)** The range of parameters $x_1 > 1$, $x_2 < 1$ marked in $(x_1, x_2)$ plane and **(b)** corresponding integral curves of equation of motion drawn for various boundary conditions. The final velocity $v_f$ is always stable and finite.

### 3.2. SOLUTIONS FOR $X_1 < 1$.

Here $\Delta$ can be either positive or negative (see Eq. 15). Thus several options are to be discussed: $\Delta < 0$, $\Delta > 0$ and $\Delta = 0$. This is what we do below:

#### 3.2.1. THE OPTION $\Delta < 0$

The domain of $(x_1, x_2)$ parameters satisfying this condition is shown in Fig.5a.
The RHS of Eq.(7) has no roots and therefore, according to (14), is positive for all $v(t) \geq 0$. It means that the block accelerates independently on its initial velocity. Our intuition indicates this possibility as the "most natural". We present integral curves for this option in Fig.5b.

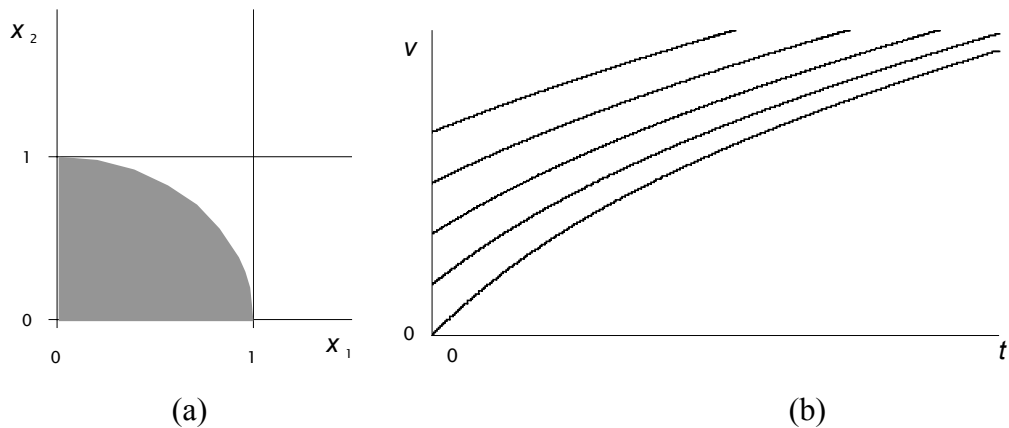

(a)          (b)

**Figure 5. (a)** The range of parameters $x_1^2 + x_2^2 < 1$ marked in $(x_1, x_2)$ plane and **(b)** corresponding integral curves of equation of motion drawn for various boundary conditions. Continuous acceleration is clearly seen.

#### 3.2.2. THE OPTION $\Delta > 0$

The area in $(x_1, x_2)$ plane corresponding to this criterion is shown in Fig.6a. The RHS of Eq.(7) is null for $v_\pm$ already given in Eq.(16). One easy checks that:

$$v_+ > v_- > 0 \qquad (21)$$

what follows from the reverse relation (17) satisfied in the considered domain ($x_1<1$, $x_2<1$), namely:

$$x_1^2 - 1 < x_2^2(x_1^2 - 1) \qquad (22)$$

Thus we have:

$$\begin{cases} \dfrac{dv}{dt} > 0, & \text{if } \quad 0 \leq v < v_- \quad \text{or} \quad v > v_+ \\ \dfrac{dv}{dt} = 0, & \text{if} \quad\quad v = v_+ \quad \text{or} \quad v = v_- \\ \dfrac{dv}{dt} < 0, & \text{if} \quad\quad\quad v_- < v < v_+ \end{cases} \qquad (23)$$

It means that $v(t) = v_-$ is the stationary and stable solution of Eq(7) while $v(t) = v_+$ is unstable, i.e. any infinitesimal increase of the initial or instant velocity above $v_+$ (or infinitesimal change in friction coefficients $f_1$, $f_2$ - what may always occur locally) leads to contrary different behaviour of the system than if this velocity is slightly below $v_+$. The situation is clarified with integral curves in Fig.6b. This diagram shows the possibility of chaotic motion of the block around $v_+$ for $x_1<1$, $x_2<1$, $x_1^2 + x_2^2 > 1$. Chaotic behaviour is usually understood as strong qualitative dependence between the solution of differential equation and the boundary condition (or parameters of considered equation) [5].

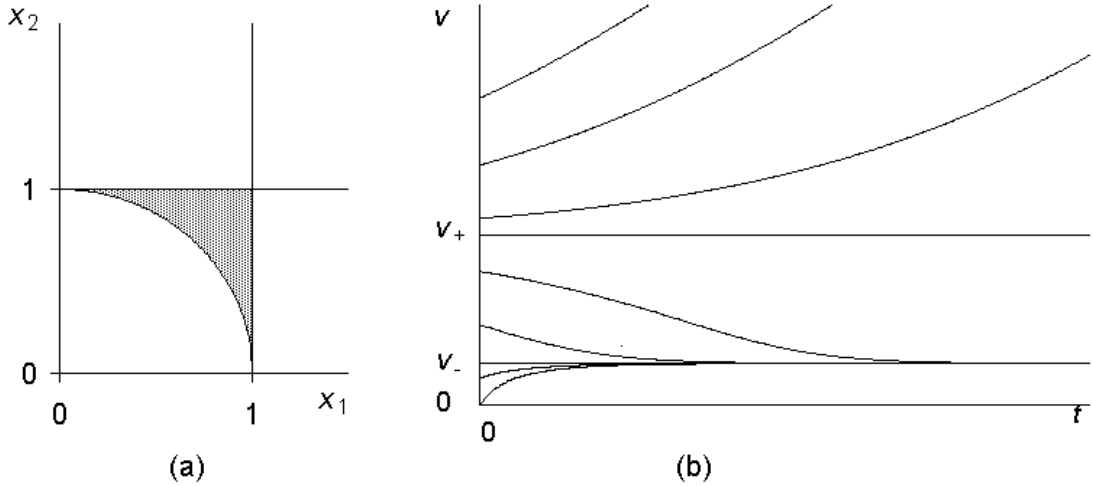

**Figure 6.** (a) The range of parameters $x_1^2 + x_2^2 > 1$, $x_1<1$, $x_2<1$ marked in ($x_1$, $x_2$) plane and (b) corresponding integral curves $v(t)$ of equation of motion drawn for various boundary conditions. Two stationary solutions $v_+(t) = $ const and $v_-(t) = $ const are seen but only one of them ($v_-(t) = $ const) is stable.

### 3.2.3. THE OPTION Δ=0
It corresponds to the constraint (see Fig.7a)

$$x_1^2 + x_2^2 = 1 \qquad (24)$$

Two roots of RHS of Eq.(7) coincide now together:

$$v_n \stackrel{df}{=} v_+ = v_- = u\frac{x_1 x_2}{1 - x_1^2} \qquad (25)$$

and make up the unstable stationary solution of differential equation as shown in Fig.7b.

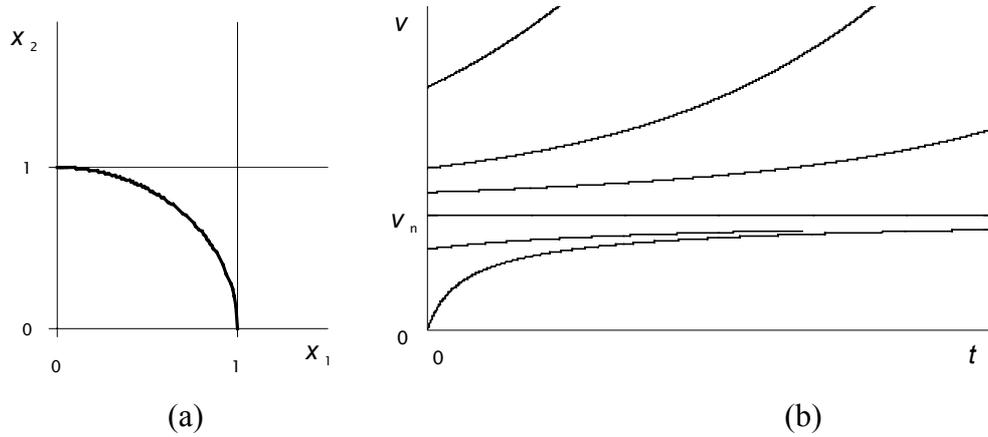

(a)                                                   (b)

**Figure 7. (a)** The range of parameters satisfying the condition $x_1^2 + x_2^2 = 1$ and **(b)** corresponding integral curves v(t) of equation of motion drawn for various boundary conditions. The unique stationary solution $v_n(t)$ = const is unstable.

### 3.3. SOLUTIONS FOR $X_1$=1.

The unique root of RHS of Eq.(7) for the above range of $(x_1, x_2)$ values is:

$$v_s = u \frac{1 - f_2^2}{2 f_2} \qquad (26)$$

It is easy to verify that:

$$\begin{cases} \dfrac{dv}{dt} > 0, & if \quad v < v_s \\ \dfrac{dv}{dt} = 0, & if \quad v = v_s \\ \dfrac{dv}{dt} < 0, & if \quad v > v_s \end{cases} \qquad (27)$$

Thus $v_s$ is the stable and stationary solution and the system evolves regularly (similarly to the case $x_1$>1) with integral curves plotted already in Fig.4b.

### 3.4. SOLUTIONS FOR $X_2 \geq 1$

Here we deal with the range of parameters shown in Figs. 8a, 9a. If $x_2 \geq 1$ the block is unable to start from the rest itself. Nevertheless the instant external action can force it to do so (see the previous section). What happens then?
If $v_0 > 0$ we are still eligible to use Eq.(7). The discriminant $\Delta$ given in Eq.(15) is obviously nonnegative for $x_2 \geq 1$ and two roots of RHS of Eq.(7) are produced, given by Eq.(16). Using the same procedure as before it is easy to verify they are both negative for $x_1 \geq 1$.

Thus the inequality (14) is never satisfied and thererefore $\dfrac{dv}{dt} < 0$ for all $v$>0 as shown in Fig. 3b.

The block will stop as visualized in Fig. 8b.

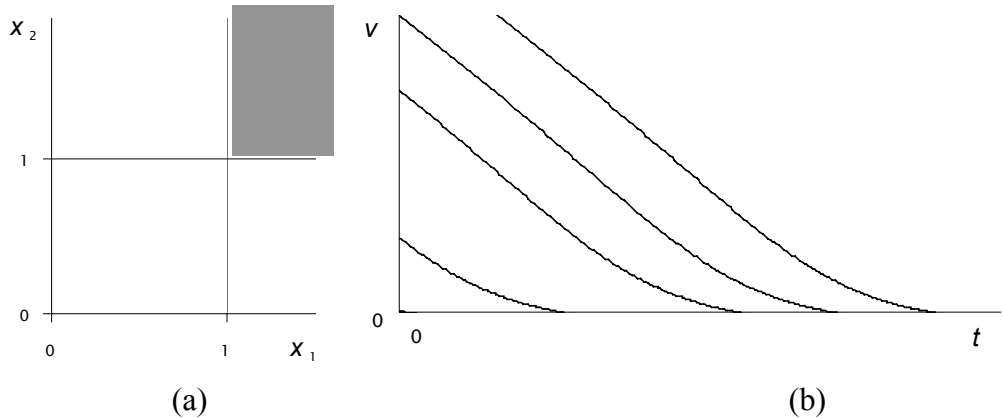

**Figure 8.** (a) The range of parameters $x_1 > 1$, $x_2 \geq 1$ marked in $(x_1, x_2)$ plane and (b) corresponding integral curves $v(t)$ of equation of motion drawn for various boundary conditions. It is seen the velocity decreases fast to zero.

Contrary, for $x_1 < 1$, relations $v_+ > 0$ and $v_- < 0$ hold ($v_- < 0$ because for $x_1 < 1$ and $x_2 > 1$ the inequality (17) is satisfied). Hence the inequality (15) also holds for any $v(t) > v_+$ so that $\frac{dv}{dt} < 0$ for $0 \leq v(t) < v_+$ and $\frac{dv}{dt} > 0$ for $v(t) > v_+$ (see Fig. 3b). It means that our block will accelerate if its initial velocity $v_0 > v_+$ and will decelerate until stopping if $v_0 < v_+$. The integral curves for this case are plotted in Fig. 9b. The behaviour around $v_+$ can be chaotic again because $v(t) = v_+$ is the stationary but unstable solution of Eq.(7).

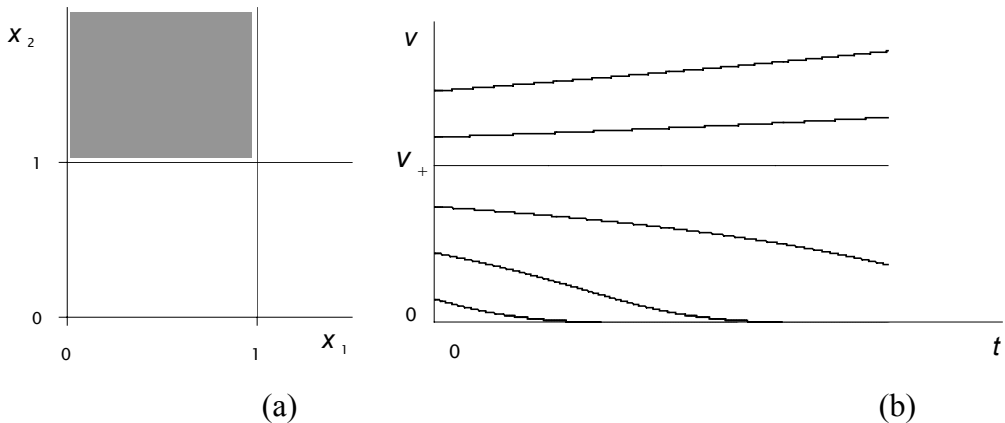

**Figure 9.** (a) The range of parameters $x_1 < 1$, $x_2 \geq 1$ marked in $(x_1, x_2)$ plane and (b) corresponding integral curves $v(t)$ of equation of motion drawn for various boundary conditions. Diagram shows instability of the unique stationary solution: $v_+(t) = $ const

## 4. CONCLUSIONS

In this article we have made a theoretical analysis of the evolution of a simple mechanical system, where the interplay between two dependent kinetic friction forces and gravity leads in same cases to interesting and unexpected results. Unexpected – in the sense that our intuition tells us the system should evolve differently than it really does. This phenomenon occurs for the particular choice of parameters $x_1$, $x_2$ describing the system and reflects the non-linearity of its differential equation of motion. In fact the shape of solutions depends only on the slope α and two coefficients of friction $f_1$, $f_2$.

The variety of obtained solutions makes considered toy-system an interesting theoretical exercise which explains how far a deep understanding of simple physical idea (e.g. kinetic friction force) is important in qualitative description of the system.

The space of physically admissible parameters $x_1 \geq 0$, $x_2 \geq 0$ may be divided into several subdomains as shown in Fig.10a. They can also be translated into ($f_1$, $f_2$) plot where they look like in Fig. 10b. Each subdomain has been analysed separately in the former section. Corresponding sets of integral curves for velocity $v(t)$ have been drawn in Figs. 4b-9b, while $\frac{dv}{dt}$ plot against $v(t)$ is shown in Figs. 3a, 3b.

The behaviour of the system is quite regular in subdomains (i), (ii), (v) (see Fig. 10). It means that $v(t)$ has the similar property independently on initial conditions (initial velocity).

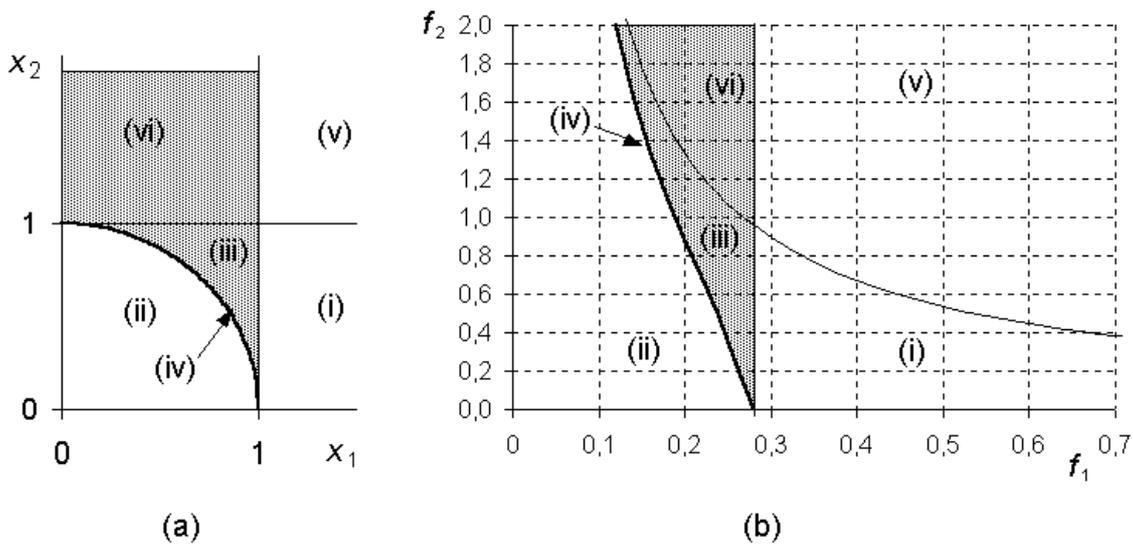

**Figure 10.** (a) The space of physically admissible parameters ($x_1$, $x_2$) divided into subdomains. The notation in subdomains corresponds to the sequence of discussion in the text.
(b) The space of physically admissible friction coefficients corresponding to these subdomains drawn for particular choice of $s$ parameter ($\alpha = \pi/12$). The hatched area corresponds to parameter range where unstable solutions of equation of motion exist.

However this situation changes drastically in remaining subdomains represented as hatched area in Fig. 10. In that case the shape of solutions and their properties (see Figs. 6b, 7b, 9b) are very sensitive to initial conditions of motion (velocity above or below $v_+$) where $v_+$ is given in Eq.(16) or in explicit form as:

$$v_+(t;\alpha,f_1,f_2) = \frac{f_1^2 f_2 \operatorname{ctg}\alpha + [f_1^2(1+f_2^2)\operatorname{ctg}^2\alpha - 1]^{1/2}}{1 - f_1^2 \operatorname{ctg}^2\alpha} u \qquad (28)$$

For $v = v_+ \pm \varepsilon$ ($\varepsilon \to 0$) there is a strong qualitative dependence of the solution from the local values of $f_1$, $f_2$ and $\alpha$ involved in $v_+$ determination. Any infinitesimal change in these parameters makes the further motion of the block qualitatively different. Since then the evolution is difficult to be predicted. It is known that such phenomenon may be the source of chaotic behaviour of the dynamical system. Basically it reflects the non-linearity of differential equation involved in the description of the process.

In the considered model this non-linearity comes from the interplay between two dependent kinetic friction forces acting in the same direction: $T_x$ and $f_2 T_y$ connected in time-dependent way according to Eqs. (3)-(5):

$$\frac{T_x}{T_y} = \operatorname{ctg}\psi(t) = \frac{v(t)}{u} \quad (29)$$

Let us finally notice that all discussed cases seem to be realistic. This is shown explicite in Fig. 10b where considered subdomains are plotted in $(f_1, f_2)$ plane as the example for $\alpha = \pi/12$. It is clear that the values of $f_1$ and $f_2$ are accessed for the wide class of materials and therefore we hope that a real model like the one considered can be built to visualise properties discussed here theoretically.

APPENDIX

One dimensional motion of mass m in OX direction under the action of net external force $F_{ext}$ and a friction force $T=fN$, where $f$ is the coefficient of kinetic friction and $N$- the normal reaction force from the surface (OX-line), is described by known formula

$$m\frac{d^2 x}{dt^2} = F_{ext} - fN \quad (A1)$$

The case discussed in this paper has however two surfaces involved. Therefore there are two different reaction forces causing friction. Two dimensional motion (in OXY reference frame) may be described by simple generalization of the formula (A1) in the following way:

$$m\frac{d^2 \vec{r}}{dt^2} = \vec{F}_{ext} + \vec{R}_\sigma + \vec{R}_W - \mathsf{F}N \quad (A2)$$

where r is a position vector of the block in OXY frame, $F_{ext}$ is the net external force (in our case played by gravity) and $R_{\sigma(W)}$ are the reaction forces from two surfaces (the plane $\sigma$ and the wall bareer W respectively).

$N = \begin{bmatrix} N_\sigma \\ N_W \end{bmatrix}$ is a 2x1 matrix with entries formed by subsequent magnitudes of reaction forces from these surfaces:

$$N_\sigma = mg\cos\alpha \quad (A3)$$
$$N_W = N_\sigma f_1 \sin\psi$$

Note that $N_\sigma \perp \sigma$ and $N_W \perp W$ so that N does not form a vector in OXY plane (as it was not in (A1) either). $\mathsf{F}$ is the generalization of friction coefficients, i.e. 2x2 friction matrix with entries depending on coefficients $f_1, f_2$ and velocity $v$.
It is easy to verify from Figs 1, 2 that

$$\vec{F}_{ext} = \begin{bmatrix} mg\sin\alpha \\ 0 \end{bmatrix}, \quad \vec{R}_\sigma = \begin{bmatrix} 0 \\ 0 \end{bmatrix}, \quad \vec{R}_W = \begin{bmatrix} 0 \\ -N_W \end{bmatrix} \quad (A4)$$

while entries of the friction matrix should be chosen as:

$$\mathsf{F} = \begin{bmatrix} f_1 \cos\psi & f_2 \\ 0 & -1 \end{bmatrix} \quad (A5)$$

to reproduce the eqation (6) as the first entry of (A2). The second entry of (A2) gives $m\frac{d^2 y}{dt^2} = 0$, and hence, from boundary conditions, $y(t)=0$.

Nondiagonal form of **F** and its dependence on velocity (via dependence on $\psi$) is the source for variety of qualitatively different solutions discussed throughout this paper. Similarity between (A2) and (A1) is the quickest alternative way to explain why the system is so mysterious. Solutions of one dimensional case (A1) would also look differently if dependence of friction coefficient $f$ on velocity $v$ was assumed.

REFERENCES:

1. Thewlis J 1962 *Encyclopaedic Dictionary of Physics* (Mac Millan, New York; Pergamon Press, Oxford) Vol. 3, p 307.
2. Serway R A 1996 *Physics for Scientists and Engineers with Modern Physics* (Saunders College Publishing) p 125.
3. Atkinson K 1993 *Elementary Numerical Analysis* (John Wiley & Sons)
4. Arnold V I 1992 *Ordinary Differential Equations* (Springer)
5. Schuster H G 1988 *Deterministic Chaos. An Introduction* (VCH Verlagsgesellschaft mbH).